\documentclass[lettersize,journal]{IEEEtran}
\PassOptionsToPackage{hyphens}{url}

\usepackage{makecell}
\usepackage{longtable}
\usepackage{booktabs}
\usepackage{float}
\usepackage{amsmath,amsfonts}
\usepackage{algorithmic}
\usepackage{algorithm}
\usepackage{array}
\usepackage[caption=false,font=normalsize,labelfont=sf,textfont=sf]{subfig}
\usepackage{textcomp}
\usepackage{stfloats}
\usepackage{url}
\usepackage{verbatim}
\usepackage{graphicx}
\usepackage{cite}
\usepackage{multirow} 
\usepackage[colorlinks,linkcolor=red,anchorcolor=green,citecolor=blue,urlcolor=black]{hyperref}
\usepackage{cleveref}

\usepackage[table]{xcolor}
\crefname{figure}{fig}{figures}
\Crefname{figure}{Fig}{Figures}
\begin{document}

\title{\emph{Wireless Copilot}: An AI-Powered Partner for Navigating Next-Generation Wireless Complexity}

\author{Haoxiang Luo, Ruichen Zhang, Yinqiu Liu, Gang Sun,~\IEEEmembership{Senior Member,~IEEE},\\Hongfang Yu,~\IEEEmembership{Senior Member,~IEEE}, and Dong In Kim,~\IEEEmembership{Life Fellow,~IEEE}
\thanks{H. Luo, G. Sun (Corresponding author), and H. Yu are with the School of Information and Communication Engineering, University of Electronic Science and Technology of China, Chengdu 611731, China (e-mail: lhx991115@163.com; \{gangsun, yuhf\}@uestc.edu.cn). R. Zhang and Y. Liu are with the College of Computing and Data Science, Nanyang Technological University, Singapore 639798 (e-mail: ruichen.zhang@ntu.edu.sg; yinqiu001@e.ntu.edu.sg). D. I. Kim is with the Department of Electrical and Computer Engineering, Sungkyunkwan University, Suwon 16419, South Korea (e-mail: dongin@skku.edu).
}

}



\maketitle

\begin{abstract}

The sixth-generation (6G) of wireless networks introduces a level of operational complexity that exceeds the limits of traditional automation and manual oversight. This paper introduces the ``\emph{Wireless Copilot}," an AI-powered technical assistant designed to function as a collaborative partner for human network designers, engineers, and operators. We posit that by integrating Large Language Models (LLMs) with a robust cognitive framework. It will interact with wireless devices, transmitting the user's intentions into the actual network execution process. Then, \emph{Wireless Copilot} can translate high-level human intent into precise, optimized, and verifiable network actions. This framework bridges the gap between human expertise and machine-scale complexity, enabling more efficient, intelligent, and trustworthy management of 6G systems. \emph{Wireless Copilot} will be a novel layer between the wireless infrastructure and the network operators. Moreover, we explore \emph{Wireless Copilot}'s methodology and analyze its application in Low-Altitude Wireless Networks (LAWNets)  assisting 6G, including network design, configuration, evaluation, and optimization. Additionally, we present a case study on intent-based LAWNets resource allocation, demonstrating its superior adaptability compared to others. Finally, we outline future directions toward creating a comprehensive human-AI collaborative ecosystem for the 6G.

\end{abstract}


\section{Introduction} \label{sec-I}

\IEEEPARstart{T}{he} 6G targets $1 \text{Tbps}$ data rates and $0.1 \text{ms}$ latency to support holographic communication and city-scale autonomous systems, 
\cite{luo2025wireless}. However, realizing this vision requires building networks of staggering complexity. 6G systems will have ultra-dense deployments of network nodes, dynamic and heterogeneous network slicing, and the seamless integration of terrestrial, aerial, and satellite networks \cite{sun2024proportional}. 
The combinatorial explosion of configurable parameters, the velocity of state changes, and the intricate interdependencies across network layers will create an operational environment. Existing rule-based systems, while effective in 4G and 5G, lack the cognitive ability to manage unforeseen scenarios in 6G \cite{chaoub2023hybrid}. This escalating complexity creates a critical need for a new management paradigm, one that is not merely automated but \emph{intelligent}, \emph{adaptive}, and \emph{interactive}.

Fortunately, concurrent breakthroughs in Artificial Intelligence (AI), particularly in the domain of Large Language Models (LLMs), offer a revolutionary path forward \cite{chen2024big}. However, simply applying AI to 6G exposes a fundamental architectural gap. Traditional machine learning approaches, while effective for specific, isolated tasks, often function as opaque ``closed boxes" \cite{luo2025ai}, making their decisions difficult to interpret, trust, or debug. These models also generalize poorly when faced with novel network scenarios not present in their training data. At the other extreme, the concept of a fully autonomous Agentic AI \cite{zhang2025toward}, which can reason, plan, and execute tasks without oversight, presents unacceptable risks for critical national infrastructure. LLMs, the core of such agents, are prone to hallucinations and unpredictable behavior \cite{luo2025weighted}, which could lead to catastrophic network configurations. 

\begin{figure}[!t]
   \centering
   \includegraphics[width=3.4 in]{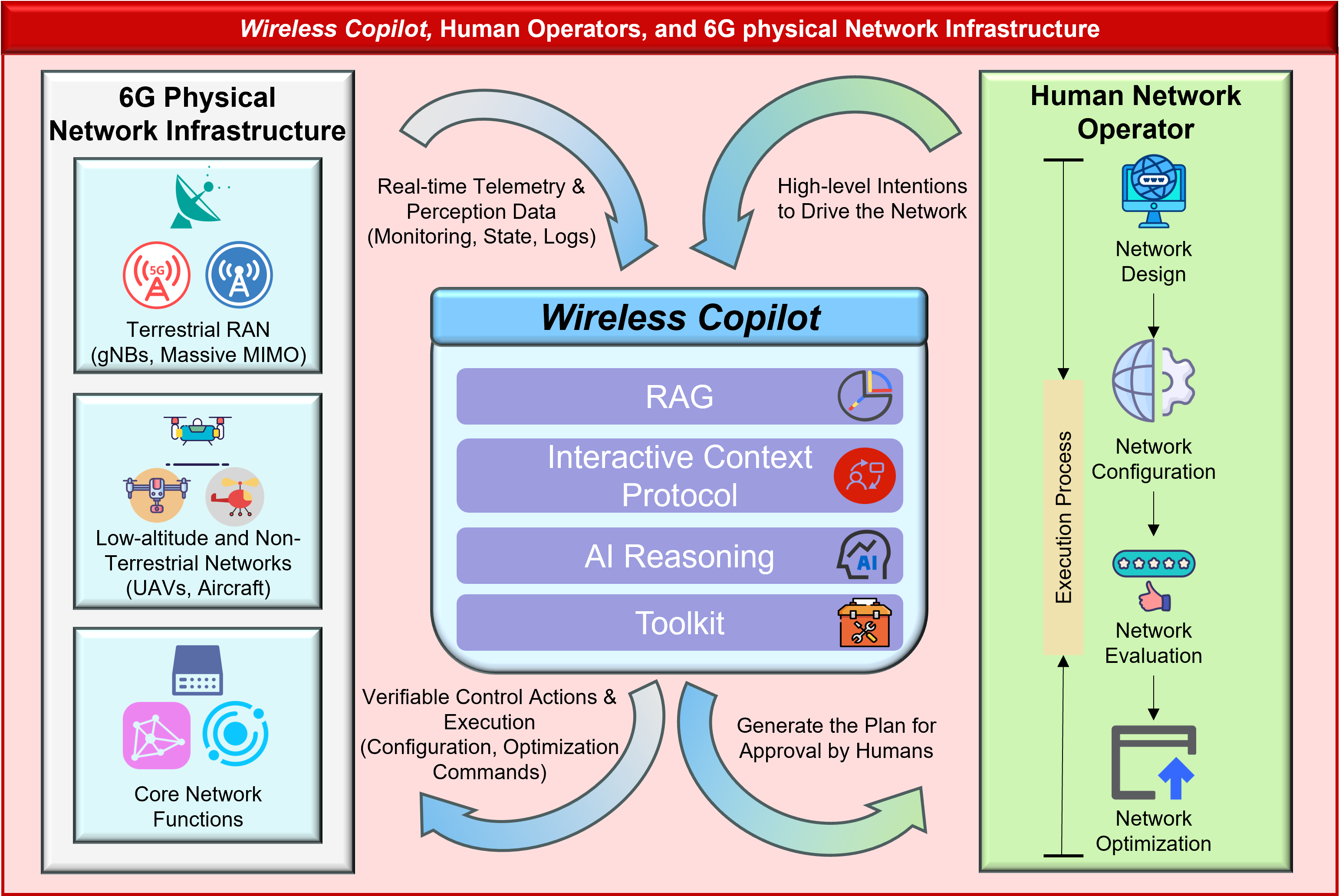}
   \caption{The interaction relationship between \emph{Wireless Copilot}, human operators, and the 6G network infrastructure. 
  \emph{Wireless Copilot} can understand the relationship and serve as a bridge between network and humans. }
   \label{fig0}
    \vspace{-0.5cm}
\end{figure}

This paper posits that the true challenge is not replacing human operators or simply bolting on AI, but designing a new cognitive management framework that enables human-AI symbiosis. 
\emph{What is missing is a framework that connects the AI reasoning capabilities to the physical network's control plane.} This framework should be human-supervised, as shown in Fig. \ref{fig0}. This gap highlights the need for an AI system that can: \emph{1) Monitor:} Ingest and understand vast streams of live network telemetry in real-time, from throughput and latency to beamforming data and routing status; 
\emph{2) Reason:} Formulate complex, multi-step operational plans based on that data and high-level human intent, using auditable processes like Chain-of-Thought (CoT);
\emph{3) Interact:} Translate human strategic goals into verifiable commands and interact directly with the physical network's control and management interfaces, such as updating Radio Access Network (RAN) parameters;
\emph{4) Validate:} Operate within a robust ``Human-in-the-Loop (HITL)" framework, where human expertise guides, validates, and retains ultimate control over all actions.

This is the motivation behind \emph{Wireless Copilot}, an AI-powered partner built upon a cognitive framework designed to augment, rather than replace, the human network operator.  Currently, AI Copilot has been applied in various scenarios. For instance, in code development, it can understand the code context and complete code segments\footnote{https://www.webmobinfo.ch/blog/ai-copilot-guide-best-tools-use-cases}. 
In contrast, \emph{Wireless Copilot} is not a simple chatbot. It is a new network management system that functions as an interactive monitoring and control layer. It translates human intent into precise, verifiable network actions. 
To the best of our knowledge, this is the first work to propose and demonstrate a cognitive framework for a \emph{Wireless Copilot} that enables direct, human-supervised interaction with the physical network. Our contributions are:

\begin{itemize}

\item We propose a cognitive framework for \emph{Wireless Copilot}. It is built on four key functions that enable a human-network interaction paradigm. This \emph{Wireless Copilot} is the novel layer between network and human operators

\item We define \emph{Wireless Copilot} as a new architectural partner for managing the network lifecycle. Using 6G-enabled Low-Altitude Wireless Networks (LAWNets), we demonstrate \emph{Wireless Copilot}'s role as a cognitive orchestrator that executes complex, cross-domain control actions.   

\item We demonstrate the effectiveness of our framework through a case study of an intent-based LAWNets resource allocation task. Our results show that \emph{Wireless Copilot}'s interactive framework achieves superior performance compared to other AI-based methods. 
\end{itemize}

\section{Preliminaries: AI Copilot} \label{sec-II}


 \subsection{Definition of AI Copilot}
 
An AI Copilot is an AI-powered virtual assistant designed to enhance user decision-making \cite{lu2024multimodal}.  
Unlike a chatbot, a Copilot possesses deep contextual awareness. It understands the user's workflow, accesses relevant data sources, and provides real-time assistance. Its primary functions include analyzing datasets to extract key insights and generating suggestions to help the user complete their work more effectively.  The defining characteristic of a Copilot is its collaborative nature. It acts as a partner, working alongside the user to navigate complexity, with the human always retaining final control.

\subsection{Why 6G Needs AI Copilot?}


The 6G management is problems that exceed human cognitive scale and defy conventional, rule-based automation.
Firstly, the extreme performance demands of 6G targets are orders of magnitude beyond those of 5G, including reliability targets as stringent as \(10^{-9}\) error rates and connection densities exceeding 10 million devices per square kilometer \cite{luo2025wireless}. Secondly, architectural heterogeneity: 6G envisions the unification of Terrestrial Networks (TNs) and Non-Terrestrial Networks (NTNs) \cite{saleh2025integrated}. This creates highly dynamic topologies, as NTN nodes are constantly in motion. Finally, intractable multi-objective optimization dilemmas: 6G network management is not about optimizing a single metric \cite{luo2025wireless}. It is a constant, high-dimensional balancing act between conflicting objectives. 

Faced with this multi-dimensional complexity, the 6G network requires an intelligence layer rooted in the HITL paradigm. One that does not merely operate autonomously, but actively reasons, adapts, and collaborates with human experts. The AI Copilot, built explicitly to embody this HITL logic, is designed to fill this role. It acts as a cognitive bridge in the HITL framework, translating human intent into optimized network reality while ensuring human expertise remains integral to every stage of decision-making and adaptation.

\subsection{Differences between LLM, Copilot, and Agentic AI}

Although both Copilot and Agentic AI rely on LLMs, and the latter has a higher level of autonomous capabilities, we still believe that future 6G will be more dependent on Copilot.


 While an Autopilot that could self-manage the network is an attractive vision, its deployment faces significant hurdles. For instance, network operations often involve high-stakes decisions with economic and safety consequences, requiring levels of strategic and ethical judgment that are currently beyond the AI capabilities. 
 The Copilot offers a pragmatic and powerful alternative. It can automate the tasks that machines are good at performing. For instance, it can sift through real-time datasets collected from 6G LAWNets, correlating spatio-temporal data points to pinpoint causes, and auto-generate configuration scripts 
 to adjust UAV beamforming parameters and flight trajectories. Simultaneously,  it will delegate those tasks that require human strategic judgment to humans. For example, in situations with limited resources, making decisions regarding the trade-off between the battery endurance and coverage range of UAVs, as well as taking ultimate responsibility for network performance, can be accomplished by human operators. 
 This HITL model fosters trust and provides a safe, evolutionary path toward greater network autonomy.

Additionally, LLM, Copilot, and Agentic AI have similar concepts, but they are different in intelligence and autonomy. 
An LLM application is primarily a reactive tool for information retrieval and text generation. A Copilot is an interactive partner that assists with tasks. An Agentic AI, or Autopilot, is a proactive system that can independently pursue goals. From the perspective of technological evolution, LLMs are key features of Copilot, and Copilot is also one of the key components of Agentic AI. Understanding these differences is crucial for deploying the right AI tool for networking.



\begin{figure*}[!t]
   \centering
   \includegraphics[width=6.7 in]{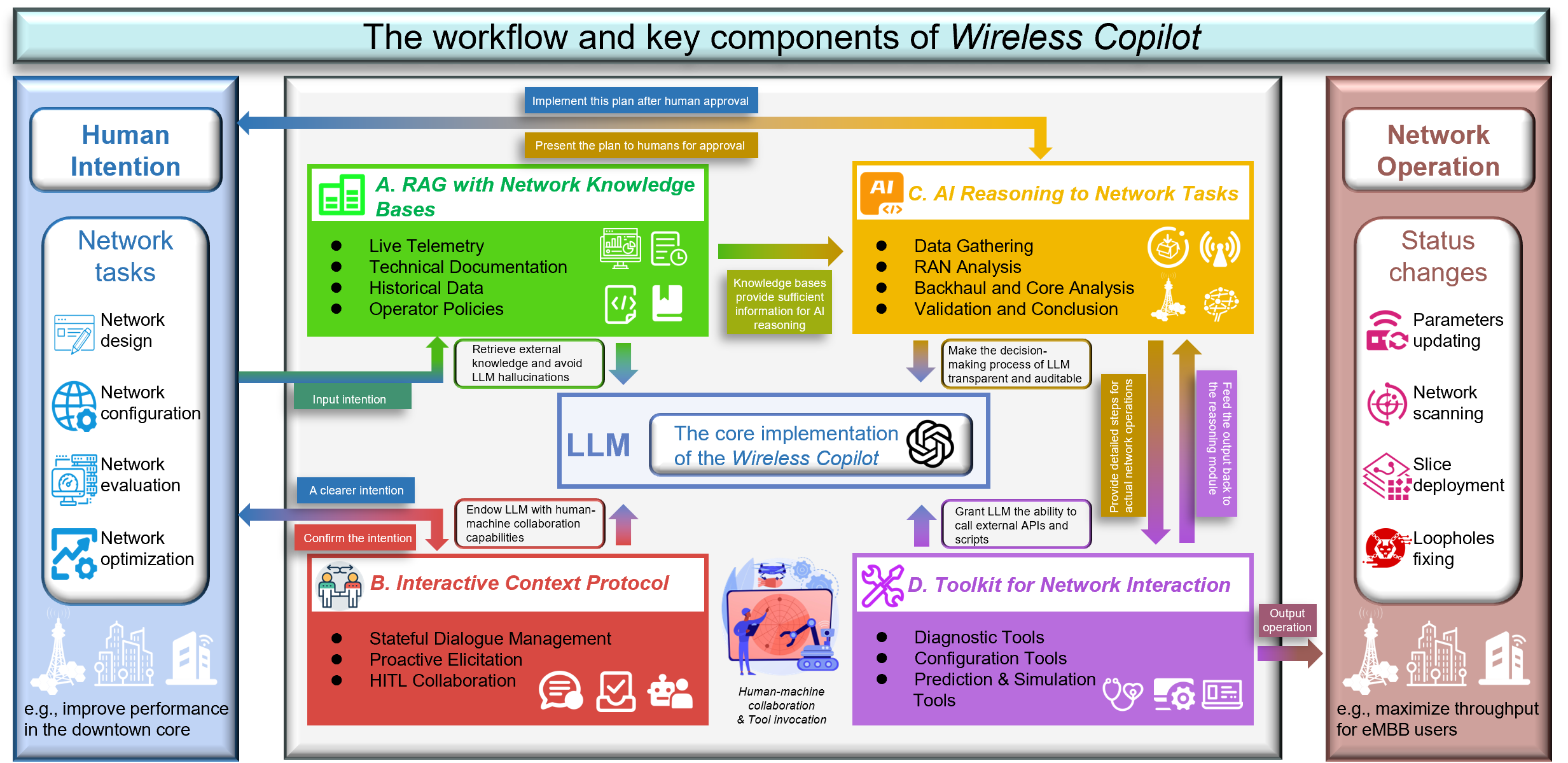}
   \caption{The workflow and key components of \emph{Wireless Copilot}. It can interact with network operators to assist humans in performing network operations.}
   \label{fig2}
    \vspace{-0.5cm}
\end{figure*}

\section{\emph{Wireless Copilot} Framework} \label{sec-III}

As shown in Fig. \ref{fig2}, it illustrates the workflow and key components of \emph{Wireless Copilot}.

\subsection{RAG with Network Knowledge Bases}


For the 6G infrastructure, hallucinations by LLM are unacceptable. RAG is the key technology that solves this problem by connecting the LLM to external and up-to-date wireless knowledge sources before it generates a response \cite{liu2025adaptive}.
For a \emph{Wireless Copilot}, this external knowledge base would be a real-time repository of all relevant network information, including:

\begin{itemize}
    \item \textbf{Live Telemetry:} Real-time performances of the 6G wireless network, e.g., throughput, latency, jitter, from across the RAN and core network. In addition to wireless-specific parameters, it also includes information such as real-time beamforming data for communication nodes and the routing protocol status of wireless links.
    
\item \textbf{Technical Documentation:} Vendor equipment manuals, 3GPP standards documents, and internal best-practice guides. It also includes technical materials such as the deployment guidelines for 6G beamforming and the optimization manual for LAWNets.

\item \textbf{Historical Data:} Logs of past network events, incidents, and their resolutions. For instance, signal coverage blind spots caused by improper configuration of beamforming parameters, or communication interruption in drone logistics due to a delay in route switching.

\item \textbf{Operator Policies:} Documents defining service level agreements (SLAs) for 6G wireless services, e.g., latency requirements for low-altitude URLLC services, wireless security policies, and resource allocation objectives.
\end{itemize}

When an operator poses a query, the RAG system first retrieves the most relevant snippets from the knowledge base and provides them to the LLM as context. This ensures that \emph{Wireless Copilot}'s plans are grounded in the actual, current state of the network and established operational principles. 

\subsection{Interactive Context Protocol}

Complex network operations cannot be managed through simple, single-turn interactions. \emph{Wireless Copilot} must engage in a multi-turn dialogue to build a shared understanding with the human operator. This requires more than just conversational ability. It necessitates a structured framework for interaction, which can be conceptualized as a context protocol. This protocol governs how the \emph{Wireless Copilot} and the operator exchange information. 
Inspired by emerging standards, including the Model Context Protocol (MCP) \cite{liu2025context}, this framework enables several critical capabilities :

\begin{itemize}
    \item \textbf{Stateful Dialogue Management:} \emph{Wireless Copilot} maintains a memory of the current conversation, including past user inputs, its own responses, and the results of any actions taken. This prevents the operator from having to repeat information and allows \emph{Wireless Copilot} to understand follow-up questions in their proper context.   

\item \textbf{Proactive Elicitation:} A key function of the protocol is to handle ambiguity. When an operator's intent is underspecified, \emph{Wireless Copilot} can proactively request clarification instead of making a risky assumption. For instance, if an operator states, ``Improve performance in the downtown core," the protocol enables the Copilot to ask for more specific information: ``Should I prioritize maximizing throughput for eMBB users or minimizing latency for URLLC services?" This interactive elicitation ensures that \emph{Wireless Copilot}'s subsequent actions are precisely aligned with the operator's true goal.

\item \textbf{HITL Collaboration:} The protocol formalizes the collaborative workflow by defining points where \emph{Wireless Copilot} must seek human approval. After formulating a plan or generating a configuration script, \emph{Wireless Copilot} can present it to the operator for review. The operator can then approve, reject, or request modifications before any action is taken on the live network. This ensures the human operator remains the ultimate authority, fostering trust and safety in critical operations.   
\end{itemize}

By standardizing the way that context is built and shared, this interactive protocol transforms the dialogue from a simple Q\&A session into a true collaborative partnership. 

\subsection{AI Reasoning to Network Tasks}

To solve problems, \emph{Wireless Copilot} must be able to formulate a logical, multi-step plan. This is achieved through AI reasoning \cite{luo2025ai}, a prominent example of which is CoT. It guides the LLM to ``think out loud" by breaking down a problem into a sequence of logical steps \cite{zhang2024chain}.
We take the intention of a network operator as an example: ``Diagnose the cause of high latency for URLLC slice users in the urban area." A \emph{Wireless Copilot} without CoT might give a general and unhelpful answer. However, a \emph{Wireless Copilot} with CoT would clearly explain its reasoning process, steps as follows:

\begin{itemize}
    \item \textbf{Data Gathering:} \emph{Wireless Copilot} retrieves the real-time latency KPIs and packet drop rates for the specified URLLC slice in the target urban geographic coordinates. Then, it aligns these data with the performance benchmarks defined in 3GPP TS 22.261\footnote{\url{https://portal.3gpp.org/desktopmodules/Specifications/SpecificationDetails.aspx?specificationId=3107}}. This standard specifies that URLLC services must meet a latency of no more than $1$ ms and a reliability of at least $10^{-9}$.

\item \textbf{RAN Analysis:} \emph{Wireless Copilot} analyzes the radio conditions for the affected URLLC users, such as signal-to-interference-plus-noise ratio (SINR). It also checks for high interference levels on the assigned physical resource blocks (PRBs) by referencing 3GPP TS 38.133\footnote{\url{https://portal.3gpp.org/desktopmodules/Specifications/SpecificationDetails.aspx?specificationId=3204}}. This standard defines the interference threshold for URLLC-specific PRBs. Additionally, it cross-validates with ITU-T G.9991\footnote{https://www.itu.int/rec/T-REC-G.9991/en}, a standard for evaluating wireless signal stability in dense urban scenarios, to rule out environmental interference factors not covered by 3GPP.

\item \textbf{Backhaul and Core Analysis:} \emph{Wireless Copilot} examines the transport network, focusing on the utilization and latency of the backhaul links connecting relevant gNodeBs to the core network. It adheres to 3GPP TS 23.501\footnote{\url{https://portal.3gpp.org/desktopmodules/Specifications/SpecificationDetails.aspx?specificationId=3144}}, which outlines QoS mapping rules for URLLC in the core network.  It also references ETSI GS NFV-IFA 014\footnote{\url{https://www.etsi.org/deliver/etsi_gs/NFV-IFA/001_099/014/03.02.01_60/gs_nfv-ifa014v030201p.pdf}}, a standard for network function virtualization (NFV) performance monitoring, to check if core network virtualized functions have latency anomalies.


\item \textbf{Validation and Conclusion:}  We set mandatory checkpoints in the CoT process. \emph{Wireless Copilot} will present the reasoning steps related to the intent to the human operator for validation, thereby forming HITL again.
By correlating these findings, \emph{Wireless Copilot} can determine the root cause. For instance, if radio conditions are good but backhaul latency is high, the problem is likely backhaul congestion.  If PRB interference exceeds the threshold in RAN analysis, the cause may be non-compliance with 3GPP radio resource management standards.
\end{itemize}

This process not only leads to a more accurate diagnosis but also makes decision-making process transparent.

\begin{figure*}[!t]
   \centering
   \includegraphics[width=6.5 in]{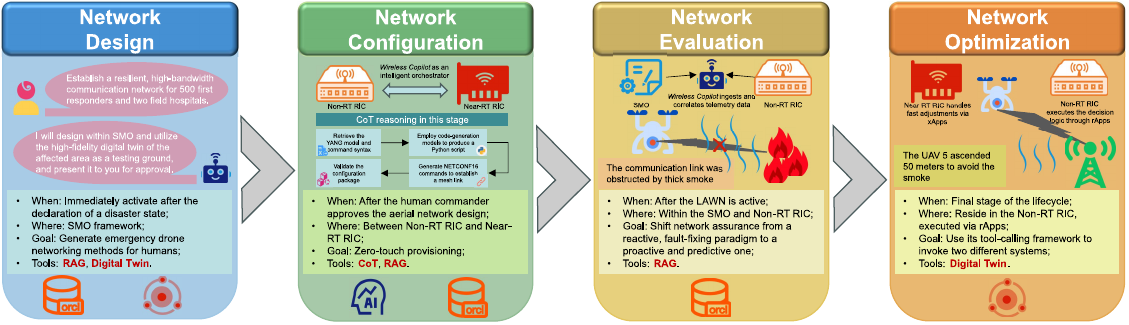}
   \caption{\emph{Wireless Copilot}'s role in transforming the network lifecycle. It can interact with network operators to assist humans in performing network operations.} 
   \label{fig3}
    \vspace{-0.5cm}
\end{figure*}

\subsection{Toolkit for Network Interaction}
To be a true partner, \emph{Wireless Copilot} must be able to act on its conclusions. The final piece of the cognitive framework is tool use, which grants the LLM the ability to invoke external software functions, scripts, or APIs. The reasoning process determines which tools to call, in what sequence, and with what parameters to execute its plan.
The toolkit for a \emph{Wireless Copilot} would be a library of functions that interface with the network's Operational Support Systems (OSS) and management frameworks \cite{qu2025llm}. This toolkit would include:

\begin{itemize}
    \item \textbf{Diagnostic Tools:} Functions such as \texttt{  run\_spectrum\_analysis(location)}, \texttt{get\_kpi\_report(slice\_id, time\_window)}, and \texttt{fetch\_event\_logs(node\_id)}.
    \item \textbf{Configuration Tools: } APIs to interact with the network orchestrator, such as \texttt{deploy\_network\_slice(config\_json)} or \texttt{update\_ran\_parameters(gNodeB\_id, params)}.
    \item \textbf{Prediction \& Simulation Tools:} Functions that call on other AI models, such as \texttt{run\_traffic\_forecast(area)} or \texttt{simulate\_beamforming\_pattern(config)}.
\end{itemize}

This combination of RAG, context protocol, reasoning, and tool use creates a powerful cognitive loop. After the operator states an intent, \emph{Wireless Copilot} uses RAG to gather necessary information.  If under-specified, it uses the context protocol to clarify ambiguities. The reasoning engine forms a plan for the operator for approval and decides to call a specific tool. The tool's output feeds back into the reasoning process, allowing \emph{Wireless Copilot} to assess results and choose next steps until the intent is fulfilled. This synergy transforms the LLM into an active, collaborative participant in network operations.

\section{Transformed 6G LAWNets by \emph{Wireless Copilot}} \label{sec-IV}

6G LAWNets provide resilience and agility for communication recovery, especially in emergency response. 
Thus, we elaborate on the transformative role of \emph{Wireless Copilot} in this scenario, as shown in Fig. \ref{fig3}.
We describe the functions that are performed by specifying the time, location, goals, and tools.

\subsection{Network Design}

\emph{Wireless Copilot} begins at the critical juncture. When human operators issue a natural language intent, such as ``Establish a resilient, high-bandwidth communication network for $500$ first responders and two field hospitals across a $20$ km$^2$ disaster zone, prioritizing QoS for medical telemetry and video drone feeds". Then \emph{Wireless Copilot}'s design function is activated.
This process unfolds entirely within the Service Management and Orchestration (SMO) framework\footnote{https://techlteworld.com/smo-service-management-and-orchestration/}, leveraging a high-fidelity Digital Twin (DT) of the disaster-stricken area as a risk-free sandbox. The primary goal is to translate this abstract intent into a multi-objective optimal architecture. 


To achieve this, \emph{Wireless Copilot}  employs a sophisticated AI toolkit. It uses RAG to ground its reasoning in reality, querying a vast knowledge base containing 3D topographical maps of the terrain, real-time weather data, and 3GPP standards for NTNs. Armed with this context, it utilizes generative AI models \cite{yao2025multi} to propose 3D placements of UAVs, which act as flying gNodeBs, forming a resilient mesh backhaul topology. Each proposed design is then rigorously stress-tested within the DT, simulating scenarios such as the dynamic movement patterns of first responder teams or the potential failure of a UAV link due to adverse weather. 
After iterative refinement, \emph{Wireless Copilot} will present the trade-offs to a human commander for final approval. It outputs a complete architectural blueprint, including UAV flight paths, gNodeB radio configurations, and a resource allocation plan. 

\subsection{Network Configuration}

Once the human commander approves the aerial network design, the lifecycle transitions to the intent-driven configuration stage. 
This phase is critical for rapid and error-free deployment in a high-pressure environment where manual configuration is untenable. \emph{Wireless Copilot} assumes the role of an intelligent orchestrator, operating between the Non-Real-Time RAN Intelligent Controller (Non-RT RIC) and the Near-Real-Time RIC (Near-RT RIC). The former defines overarching policies like slice QoS parameters based on the design intent, and the latter receives dynamic control parameters for execution. 

The core objective here is to achieve zero-touch provisioning, accurately translating the high-level design blueprint into lines of device-specific configurations. Thereby, it can eliminate the risk of human error and configuration drift, which are the primary causes of network failure. 

\emph{Wireless Copilot}’s methodology is anchored in CoT reasoning, which allows it to deconstruct a complex intent into a logical, auditable sequence of actions. For instance, the intent ``Configure UAV-5 as the primary backhaul node for Field Hospital 1" initiates a CoT process:
\begin{enumerate}
    \item Use RAG to retrieve the specific YANG model and command syntax for UAV-5's onboard gNodeB. 
    \item Employ specialized code-generation models to produce a Python script that configures radio parameters and establishes QoS queues prioritizing medical telemetry traffic with a guaranteed bit rate. 
   \item Generate NETCONF commands to establish a secure, high-capacity mesh link to designated relay UAVs. 
  \item Validate the complete configuration package in the DT to verify end-to-end connectivity and latency.
\end{enumerate}
Upon successful virtual validation, \emph{Wireless Copilot} uses its tool-use capabilities to invoke network automation APIs. For instance, standardized O-RAN interfaces such as O1 and A1\footnote{\url{https://www.rfwireless-world.com/terminology/o-ran-architecture-interfaces}} can communicate with the ground control system of the UAVs. Then, it pushes the verified configurations to the live UAVs and brings the emergency network online.   

\subsection{Network Evaluation}
From the moment the LAWNet is active, \emph{Wireless Copilot}'s evaluation and assurance function operates continuously, providing 24/7 oversight. This stage is not merely reactive. While it can be triggered for post-incident root cause analysis, its primary function is proactive and can forecast potential service degradations \cite{chen2024big}. 
Meanwhile, the main analysis engine resides within the SMO and Non-RT RIC, where it ingests and correlates vast streams of telemetry data from every UAV, network function, and connected device. For time-sensitive issues, lightweight anomaly detection agents can be deployed to the Near-RT RIC for faster, localized analysis. 

The overarching goal is to fundamentally shift network assurance from a reactive, fault-fixing paradigm to a proactive and predictive one. This involves identifying potential issues before they impact mission-critical services, such as the responder's communication link. For example, the QoE for a high-definition video feed from a field hospital begins to degrade. A legacy system might only raise a generic ``packet loss" alarm. \emph{Wireless Copilot}, however, employs multi-modal data fusion. It correlates the QoE drop with a slight increase in packet loss on UAV-5's backhaul link but rules out radio frequency interference or network congestion by analyzing physical layer data. Critically, it ingests and analyzes the video feed from UAV-5's own navigational camera. This data reveals the UAV is flying through thick smoke from a nearby fire.

 Furthermore, using RAG to consult its knowledge base on mmWave propagation characteristics \cite{luo2025wireless}, \emph{Wireless Copilot} generates a grounded diagnosis: ``Root cause is due to atmospheric absorption and scattering of the mmWave signal by dense particulates." This level of diagnostic precision is impossible with traditional tools and is essential for triggering an effective optimization response.
 
\begin{figure*}[!t]
   \centering
   \includegraphics[width=6 in]{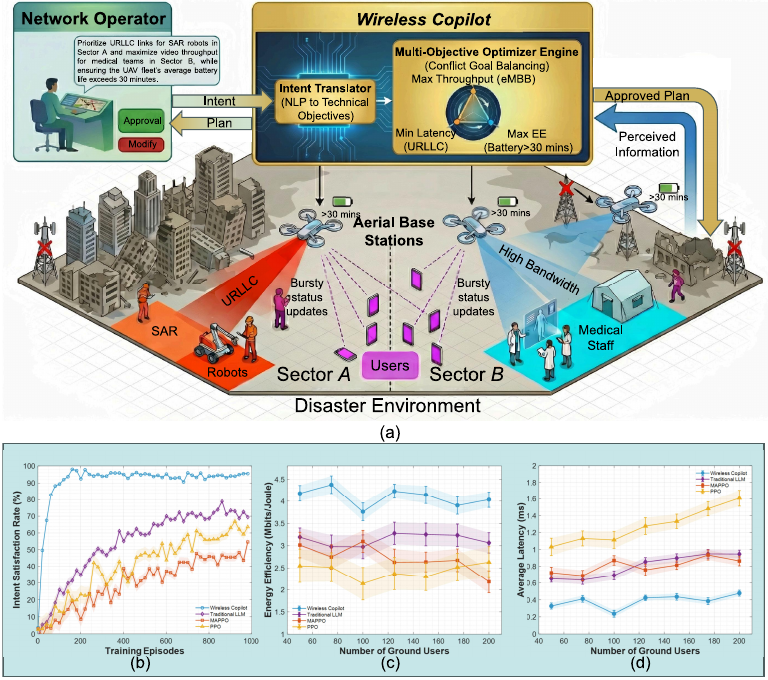}
   \caption{(a) Case scenario: Intent-driven resource allocation for 6G LAWNets in post-seismic disaster response; (b) Intent Satisfaction Rate (ISR) convergence; (c) Energy Efficiency (EE); (d) Average latency violation rate.} 
   \label{fig5}
    \vspace{-0.5cm}
\end{figure*}

\subsection{Network Optimization}

The final stage of the lifecycle, autonomous optimization, is triggered by the precise diagnosis delivered by the evaluation engine, initiating a self-healing action. It can also be triggered proactively by predictive models, leading to self-optimization. This creates a fully autonomous, closed-loop control system that ensures the network's operational state remains continuously aligned with the original mission intent. The important point is that the network can adapt in real time and dynamically to disaster situations. 

The decision-making logic for network-wide optimizations resides in the Non-RT RIC, executed via rApps. While the Near-RT RIC handles fast adjustments, such as beamforming, via xApps. Following the diagnosis of ``signal absorption due to smoke," a high-priority optimization intent is triggered. 

An LLM agent within the Non-RT RIC, extensively trained on millions of scenarios within the DT \cite{zhang2025toward}. It considers increasing the link's transmit power, re-routing traffic through a more distant UAV, or commanding UAV-5 to change its physical location. The LLM agent concludes that the optimal strategy is to have UAV-5 increase its altitude by 50 meters, flying above the smoke plume to re-establish a clear Line-of-Sight (LoS) link. 

\emph{Wireless Copilot} uses its tool-calling framework to invoke two different systems simultaneously. It sends a command to the UAV's ground control API to execute the altitude change \texttt{set\_flight\_altitude(uav\_5, new\_alt)}, while concurrently instructing an xApp in the Near-RT RIC to dynamically manage the beamforming patterns to maintain link stability during the ascent. This seamless, closed-loop action transitions from data correlation to diagnosis to cross-domain physical actuation efficiently. 


\section{Case Study} \label{sec-V}
\subsection{Case Scenario and Simulation Setup}
To demonstrate the capabilities of \emph{Wireless Copilot}, we present a mission-critical disaster response scenario. As shown in Fig. \ref{fig5} (a), following a seismic event in a city center, terrestrial base stations are compromised. A fleet of autonomous UAVs, acting as aerial base stations, is deployed to restore 6G connectivity for three distinct user groups: Search-And-Rescue (SAR) teams requiring URLLC for rescue robots, medical staff needing high-bandwidth holographic consultations, and trapped civilians sending bursty status updates. Then, the network operator issues a natural language intent: ``Prioritize URLLC links for SAR robots in Sector \emph{A} and maximize video throughput for medical teams in Sector \emph{B}, while ensuring the UAV fleet's average battery life exceeds 30 minutes". 
\emph{Wireless Copilot} need translate this intent into specific flight trajectories and radio resource allocations in real-time.

The simulation is in a $500 \times 500$ m area. We deploy $4$ rotary-wing UAVs serving ground users distributed via a Gauss-Markov mobility model \cite{ji2024resource}. The communication links operate at a carrier frequency of $28$ GHz with a bandwidth of $400$ MHz. The channel follows the 3GPP TR 38.901 Urban Micro model\footnote{\url{https://portal.3gpp.org/desktopmodules/Specifications/SpecificationDetails.aspx?specificationId=3173}}, explicitly accounting for blockage with a LoS path loss exponent of $2.1$ and a Non-LoS exponent of $3.5$. UAV power consumption is modeled using Blade Element Momentum theory, including blade profile power $79.8$ W, induced power $88.6$ W, and rotor tip speed $120$ m/s.  Moreover, we compare \emph{Wireless Copilot} with an LLM-based scheme, Multi-Agent Proximal Policy Optimization (MAPPO) \cite{yao2025multi}, and Proximal Policy Optimization (PPO) \cite{sun2024proportional}. 
The LLM-based scheme and \emph{Wireless Copilot} are developed by ChatGPT-4o. 

\subsection{Simulation Results}

Fig. \ref{fig5} (b) compares Intent Satisfaction Rate (ISR) across the four schemes, which is defined as the percentage of time slots that QoS constraints specified in the intent are satisfied.  \emph{Wireless Copilot} leads with $94.2\%$ ISR and $<50$ episodes of convergence, followed by Traditional LLM, MAPPO, and PPO. The gap hinges on HITL integration. Unlike traditional LLM, which only performs algorithm-level dynamic rule adjustment, \emph{Wireless Copilot} embeds human real-time intent calibration and domain knowledge pruning into the LLM. It can avoid semantic and QoS mismatches, as well as trial-and-error. Traditional LLM lacks this human feedback, leading to misaligned dynamic rules and lower ISR. MAPPO and PPO, without LLM and HITL, struggle with high-dimensional state space exploration or local throughput bias.
Fig. \ref{fig5} (c) shows the Energy Efficiency (EE) of three schemes. \emph{Wireless Copilot} maintains the most stable and highest EE of 4.1 Mbits/Joule. The key is also HITL-driven domain knowledge injection. \emph{Wireless Copilot}’s LLM not only dynamically optimizes UAV protocols but also integrates human empirical energy insights via HITL, avoiding energy traps. Traditional LLMs’ dynamic adjustments lack human calibration, deviating from actual UAV energy characteristics. MAPPO and PPO suffer from coordination overhead or redundant movements
In Fig. \ref{fig5} (d), \emph{Wireless Copilot} achieves the lowest average latency and the smoothest growth trend, with values ranging from 0.3 to 0.5 ms. This is because \emph{Wireless Copilot} uses human real-time feedback to refine traffic priority rules, 
alongside RL resource optimization. Traditional LLM-driven classification lacks human calibration, leading to misjudgment and resource misallocation. MAPPO and PPO lack priority awareness, causing latency-sensitive traffic resource starvation.

\section{Conclusion and Future Directions}
\label{sec-VI}
In this paper, we propose \emph{Wireless Copilot}, an AI-powered collaborative framework for managing the operational complexity of 6G networks. Integrating retrieval-augmented generation, interactive context protocols, AI reasoning, and network toolkits, the framework realizes human-AI symbiosis and bridges the gap between human expertise and machine-scale network complexity. We demonstrate its transformative value across the full lifecycle of 6G low-altitude wireless networks, covering network design, configuration, evaluation, and optimization, and validate its superior performance in multi-objective resource allocation through a case study. 

Looking forward, three core research directions will be pursued to enhance the practical applicability of the framework: \emph{1) dynamic intent understanding and multi-stakeholder collaboration}, enabling intent fusion, conflict resolution, and collaborative decision-making for multi-party interactions; \emph{2)  a unified cross-domain tool ecosystem with standardized APIs} to support seamless interaction across terrestrial, aerial, and satellite heterogeneous networks; \emph{3) real-time robust reasoning and edge deployment}, optimized via model compression, edge-native fine-tuning, and lightweight algorithms to satisfy the ultra-low latency constraints of 6G systems.

\bibliographystyle{IEEEtran}
\bibliography{IEEEabrv,mylib}



\vspace{3em}

\vfill

\end{document}